# Catalysis by metallic nanoparticles in solution: Thermosensitive microgels as nanoreactors

*Rafael Roa[1,§], Stefano Angioletti-Uberti,[2,3], Yan Lu[1], Joachim Dzubiella[1,4], Francesco Piazza[5], Matthias Ballauff[1,4]*

[1] Soft Matter and Functional Materials, Helmholtz-Zentrum Berlin für Materialien und Energie GmbH, 14109 Berlin, Germany

[2]Department of Materials, Imperial College London, London SW72AZ, UK

[3]Beijing Advanced Innovation Centre for Soft Matter Science and Engineering, Beijing University of Chemical Technology, 100099 Beijing, PR China

[4]Institut für Physik, Humboldt-Universität zu Berlin, 12489 Berlin, Germany

[5]Université d'Orléans, Centre de Biophysique Moléculaire, CNRS-UPR4301, 45071 Orléans, France

*Dedicated to Prof. E. Rühl on the occasion of his 60. Birthday*

[§]*present address*: Física Aplicada I, Universidad de Málaga, 29071 Málaga, Spain

KEYWORDS: catalysis, metal nanoparticles, nanoreactor, reduction, thermosensitive network

**Abstract**: Metallic nanoparticles have been used as catalysts for various reactions, and the huge literature on the subject is hard to overlook. In many applications, the nanoparticles must be affixed to a colloidal carrier for easy handling during catalysis. These "passive carriers" (e.g., dendrimers) serve for a controlled synthesis of the nanoparticles and prevent coagulation during catalysis. Recently, hybrids from nanoparticles and polymers have been developed that allow us to change the catalytic activity of the nanoparticles by external triggers. In particular, single nanoparticles embedded in a thermosensitive network made from poly(N-isopropylacrylamide) (PNIPAM) have become the most-studied examples of such hybrids: Immersed in cold water, the PNIPAM network is hydrophilic and fully swollen. In this state, hydrophilic substrates can diffuse easily through the network, and react at the surface of the nanoparticles. Above the volume transition located at 32°C, the network becomes hydrophobic and shrinks. Now hydrophobic substrates will preferably diffuse through the network and react with other substrates in the reaction catalyzed by the enclosed nanoparticle. Such "active carriers", may thus be viewed as true nanoreactors that open new ways for the use of nanoparticles in catalysis.

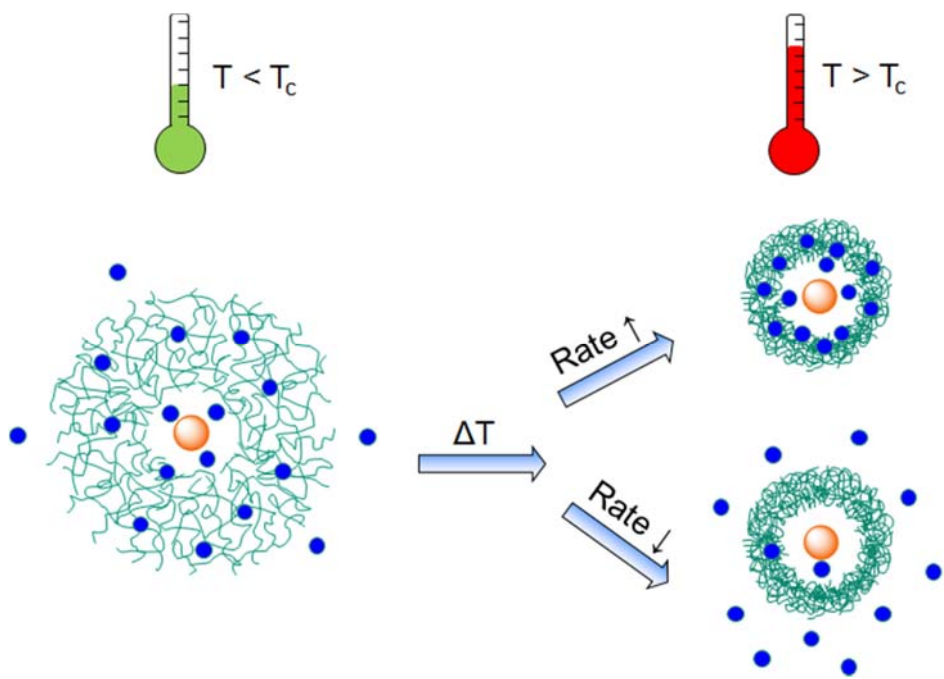

In this review, we give a survey on recent work done on these hybrids and their application in catalysis. The aim of this review is threefold: We first review hybrid systems composed of nanoparticles and thermosensitive networks and compare these "active carriers" to other colloidal and polymeric carriers (e.g., dendrimers). In a second step we discuss the model reactions used to obtain precise kinetic data on the catalytic activity of nanoparticles in various carriers and environments. These kinetic data allow us to present a fully quantitative comparison of different nanoreactors. In a final section we shall present the salient points of recent efforts in the theoretical modeling of these nanoreactors. By accounting for the presence of a free-energy landscape for the reactants' diffusive approach towards the catalytic nanoparticle, arising from solvent-reactant and polymeric shell-reactant interactions, these models are capable of explaining the emergence of all the important features observed so far in studies of nanoreactors. The present survey also suggests that such models may be used for the design of future carrier systems adapted to a given reaction and solvent.

## 1. INTRODUCTION

Metallic nanoparticles[1–4] and nanoalloys[5] have been the subject of intense research in the last two decades because of their catalytic activity. For example, gold becomes an active catalyst when divided down to the nanophase.[6–8] This discovery has led to intense research in the field and the number of papers in this subject is hard to overlook.[1,9,10] It is hence fair to state that catalysis by nanoparticles is among the most active fields in modern nanoscience. Very often the synthesis of nanoparticles proceeds through wet-chemical methods in the aqueous phase. The precipitation or co-precipitation of metal nanoparticles and alloys in water can be handled in a well-controlled manner to yield particles with defined size and composition.[1] It is therefore advantageous to use the aqueous suspension of the nanoparticles directly for catalysis. Problems of solubility of organic compounds in water can be circumvented in many cases by working in a two-phase system as, e.g., in case of the Heck- or the Suzuki reaction.

A problem that has accompanied the field from its early beginnings, however, is the handling of the particles in the liquid phase: The surface of the particles should be easily accessible for the mixture of the reactants. This condition would require the nanoparticles to be freely suspended in the solution. On the other hand, the nanoparticles must be removed entirely from solution after catalysis. Moreover, coagulation or any type of Ostwald ripening of the nanoparticles must not occur during the catalytic reaction. Also, leaching of metal or loss of nanoparticles from the carrier must be prevented to ensure a meaningful and repeated use of the catalyst. The latter requirements which are fundamental, for catalysis necessitate a suitable carrier that ensures a safe and repeated handling of the nanoparticles that may impose possible health hazards.[11,12]

Colloidal and polymeric carrier systems provide a solution to these problems. Suitable carrier systems include colloidal particles[13,14], dendrimers[15–26] mesoporous materials,[4,27,28] spherical polyelectrolyte brushes (SPB)[29,30] and other systems[31] structured on a length scale between one and a few hundred nm. At first, these systems may be used to generate the nanoparticles in a defined way. Thus, these systems enormously widen the synthetic pathways to nanoparticles. In addition to this, these carriers allow us to handle particles in solution in a secure way. Figure 1a displays such systems in a schematic fashion. In the following, these systems will be referred to as "*passive carriers*".

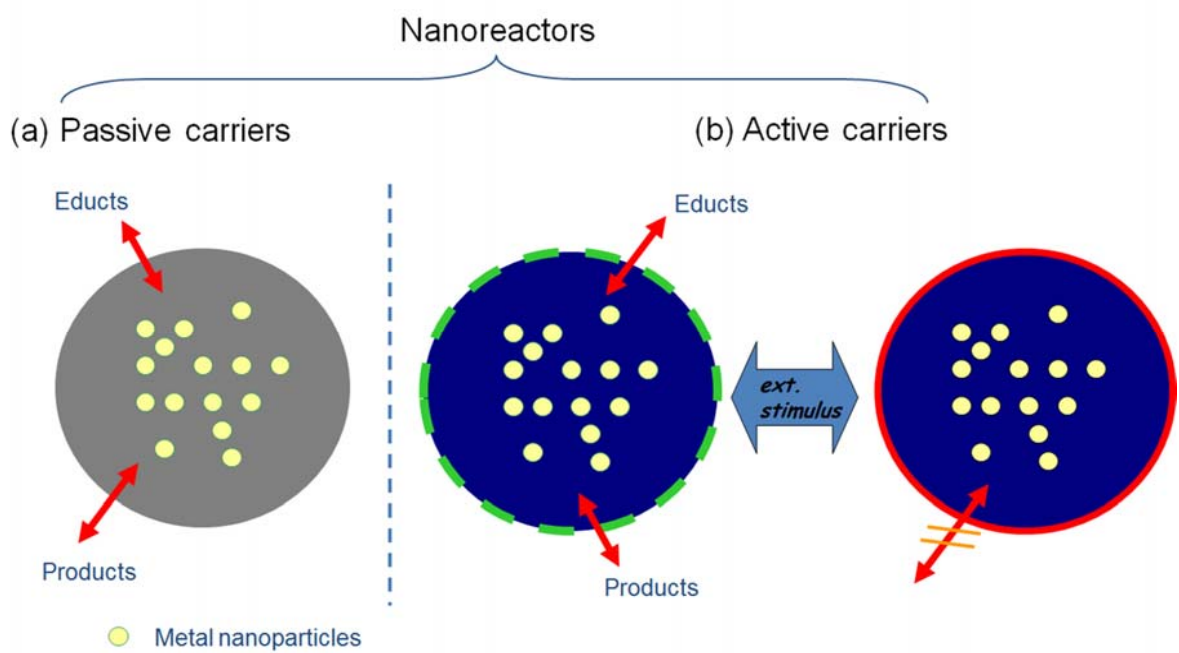

**Figure 1**: Scheme of a nanoreactor. a) *Passive systems* are carriers (e.g., dendrimers or spherical polyelectrolyte brushes) that just confine the nanoparticles and prevent their coagulation during catalysis. Reactants and products (yellow beads) can enter and exit the carrier easily. b) *Active carriers, that is, true nanoreactors* can be used to trigger the activity of the nanoparticles by an external stimulus, for example temperature or pH. Thus, the external stimulus may deplete a given substrate, hence to a certain extent the nanoreactors is "closed" for this reactant. Similarly, substrates may be enriched within the nanoreactor leading to a higher reaction rate.

In recent years, the concept of such carrier systems has been further advanced with the synthesis of "*nanoreactors*".[32–44] As depicted schematically in Figure 1b, a true nanoreactor should not only allow us to generate and to handle metallic or oxidic nanoparticles in the liquid phase without loss or leaching of material. It should be possible to change the catalytic activity and the catalytic selectivity of nanoparticle by an external trigger, for example the system's temperature or pH. These systems will be henceforth named "*active carriers*" and will be discussed in detail further below.

Thermosensitive colloidal microgels made from a network of poly(N-isopropylacrylamide) (PNIPAM) and its copolymers provide a good example for such nanoreactors:[36,37,45–53] Metallic nanoparticles are fixated within a thermosensitive network. Immersed in cold water the network will shrink. Above the temperature of the volume transition, which is 32°C for PNIPAM, most of the water will be expelled. The substrates and the products can diffuse with only little obstruction through the network in the swollen state. In the shrunken state, however, the much denser network imposes a higher diffusional resistance for the substrates and the products. The activity of the nanoparticles immobilized within such a thermosensitive network can hence be changed by temperature, as shown in a number of papers.[32,46,54–60] In our recent work, we could demonstrate that polymeric nanoreactors can also be used to tune not only the activity but also the selectivity of catalysis.[36] We showed this by synthesizing Au-nanoparticles enclosed in a hollow polymeric shell in aqueous solution, leading to a yolk-shell type of architecture. The catalytic activity of such a nanoreactor towards substrates differing in hydrophilicity can be tuned by temperature: At low temperature the PNIPAM network is hydrophilic and hydrophilic substrates will penetrate the network and react preferably. At higher temperatures the network becomes more hydrophobic which leads to a higher reactivity of hydrophobic substrates. Another example has been presented by Horecha et al.[61], who observed much higher reaction rates when Ag nanoparticles on SiO$_2$ were encapsulated in a microcapsule as compared to the activity of the free nanoparticles. These very promising results open the way for a new route towards adjusting the reactivity and the selectivity of catalytic nanoparticles. More recently, we have extended the "nanoreactor" concept to photocatalysis.[62] We have found that Cu$_2$O@PNIPAM core-shell microgels can work efficiently as photocatalyst for the decomposition of methyl orange under visible light. A significant enhancement in the catalytic activity has been observed for the core-shell microgels compared with the pure Cu$_2$O nanocubes. Most importantly, the photocatalytic activity of the Cu$_2$O nanocubes can be further tuned by the thermosensitive PNIPAM shell.[62]

A quantitative study of a nanoreactor requires kinetic data measured with the highest precision possible. Up to now, most of the testing of the catalytic activity of nanoparticles in aqueous phase has been done using the reduction of 4-nitrophenol by borohydride. Pal et al.[63] and Esumi et al.[19] have been the first who have demonstrated the usefulness of this reaction. In the meantime, the reduction of p-nitrophenol has become the most used model reaction for the quantitative testing and analyzing of the catalytic activity of nanoparticles in the liquid phase.[64,65] Precise kinetic data obtained by such a suitable model reaction can then be used to compare different metal nanoparticles immobilized in different carrier systems in a quantitative fashion. Up to now, there is an abundant literature devoted to studies in which different carrier systems are compared on this basis.

Compared to the abundant experimental literature on nanoreactors, the number of papers devoted to their theory is scarce. Let us consider a catalytic nanoparticle embedded in a polymer network. At first, diffusional control comes into play which may be treated with well-known concepts of reaction kinetics.[66] Carregal-Romero et al.[55] have been the first to model the catalytic activity of a single gold nanoparticle enclosed in a thermosensitive PNIPAM-network. These authors treated the slowing down of the rate with the shrinking of the network in terms of a decreased diffusion coefficient in the collapsed network. Recently, we have presented an extended theory of diffusion-limited reactions in nanoreactors, which can be employed to rationally design the activity and selectivity of a nanoreactor.[67,68] We called attention to the fact that a thermosensitive network may solubilize or repel reactants depending on their polarity. Mass transport in a nanoreactor must hence be described as diffusion through a free-energy landscape which also determines the local reactant concentration, i.e, partitioning inside the network. In this respect, we define the nanoreactor shell permeability as the product of the reactant partitioning and diffusivity.[68–70] Permeability can be thus defined as the inverse of a diffusional resistance of a medium regarding the total mass transport (flux) towards the catalyst driven by the reaction. Thus, quantitative concepts have emerged recently that will lead to fully quantitative understanding of nanoreactors in the future.

Here we review recent work done on nanoreactors that are built on thermosensitive networks. These active carrier systems are compared to passive carrier systems on the basis of quantitative kinetic data. The review is organized as follows: In the following section 2 a brief survey of carrier systems used in aqueous phase so far will be given. Section 3 is devoted to a discussion of model reactions used for testing the catalytic activity of nanoparticles in aqueous phase so far. Here special emphasis is laid on the reduction of 4-nitrophenol for which a huge set of data has now become available.[64,65] Section 4 presents a survey of theoretical models that describe the interplay between diffusion and chemical reactivity in such a nanoreactor.[67] A short conclusion will wrap up the entire discussion.

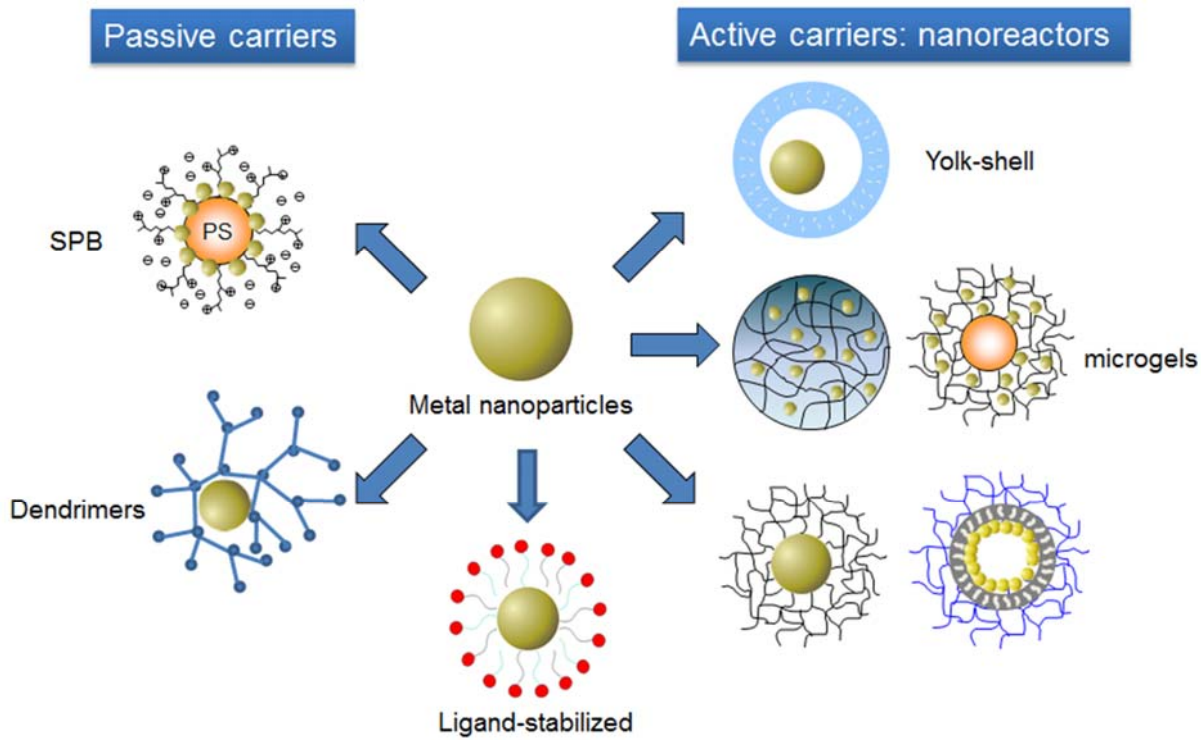

**Figure 2.** Schematic survey of the carrier systems for the immobilization of metal nanoparticles. On the left-hand side, dendrimers and spherical polyelectrolyte brushes (SPB) present examples of passive carriers that can be used for the synthesis and secure handling of the nanoparticles in catalysis. The right-hand side displays a survey of active nanoreactors (e.g., polymeric microgels and yolk-shell systems). These systems allow us to enhance or reduce the catalytic activity of the embedded nanoparticles by an external trigger. Free nanoparticles stabilized by charges or small ligands, on the other hand, may be used as reference systems in catalytic studies comparing kinetic data of nanoparticles embedded in different carrier systems.

## 2. Carriers and Nanoreactors: A survey

Figure 2 gives an overview of a number of carrier systems used so far. As mentioned above, there are passive carrier systems that are classical colloidal methods to stabilize nanoparticles in solution. In addition, there are active carrier systems that can be used as nanoreactor.

*Passive carrier systems*: First of all, nanoparticles can be stabilized in solution by suitable surface modification, for example through ionic charges. Organic ligands have been also often used to stabilize metal nanoparticles. However, it has been demonstrated that ligand stabilization could decrease the catalytic activity of nanoparticles in some cases.[71–73] Thus, it is crucial to choose an appropriate capping ligand, which can enhance both stability and activity of the particles. For instance, Astruc et al.[74] have studied the stereoelectronic effects of the ligands on the catalytic activity of the gold nanoparticles varying from ionic thiolate or citrate ligands to neutral triazole ligands.

Dendrimers can stabilize metal nanoparticles by their functional groups that coordinate to the metal nanoparticle surface with minimal impedance of catalytic activity.[15–26,75–80] The most common used dendrimers are the commercially available poly(amidoamine) (PAMAM) type dendrimers.[17,81–83] Esumi et al.[19] have synthesized Pt, Pd and Ag nanoparticles using PAMAM dendrimers as stabilizer. The influence of various generations of dendrimers on the catalytic activity of the metal nanoparticles has been investigated. Recently, Meijboom et al.[15,20–24] performed a number of careful kinetic studies on the catalytic activity of dendrimer-encapsulated nanoparticles. Astruc et al. have demonstrated that the catalytic activities of Ag, Au, and Cu nanoparticles stabilized by dendrimers using coordination to intra-dendritic triazoles, galvanic replacement or stabilization outside dendrimers strongly depend on their location.[84]

Polymer micelles and microemulsions with functional groups that can complex with metal ions or precursors have been widely applied as carrier systems for the generation of metal nanoparticles.[85–87] For example, polymeric micelles from various amphiphilic block copolymers, such as polystyrene-*b*-poly-4-vinylpyridine (PS-*b*-P4VP), have been synthesized and used for the immobilization of metal nanoparticles.[88–90]

Some time ago we demonstrated that spherical polyelectrolyte brushes (SPB)[29,30] are highly suitable carrier system for the immobilization of various metal nanoparticles.[13,14,38,91–96] The SPB particles consist of a polystyrene core onto which a dense layer of polyelectrolyte brushes is grafted. The metal ions are immobilized in the brush layer as counter-ions. Reduction of these immobilized metal ions with $NaBH_4$ leads to nanoparticles of the respective metal.[13,14] Various metal nanoparticles (Au, Pd and Pt) or nanoalloys (Au/Pt, Au/Pd) have been anchored on the SPB particles in this way.[97]

*Active carriers and nanoreactors*: Networks from thermosensitive polymers, in particular made from crosslinked poly(N-isopropyl-acrylamide) (PNIPA) have been extensively used for the assembly of colloidal hybrids[98] and nanoreactors.[99] Dispersed in cold water, such a network will swell while most of the water will be expelled above the temperature of the volume phase transition, which is 32°C for PNIPAM. Thus, PNIPAM microgels or core-shell microgels based on PNIPAM have been developed as active nanoreactor for the immobilization of different metal nanoparticles.[34,45,46,55,100–102] In these systems the dependence of the catalytic activity of the embedded nanoparticles can be changed by the temperature, which acts as an external trigger. Comonomers can be built into the network to provide an additional sensitivity towards other triggers, for example the pH in the system. Recently, Q. Wu et al. presented a microgel that can be shrunken or swollen by adding glucose to the solution which in turn modified the chemical reaction.[52] Li et al. developed a new type of smart composite microgels, which are able to control the catalytic activity of their loaded silver nanoparticles by light, by immobilization of Ag nanoparticles into the shell of the AuNR@PNIPAM core-shell microgels containing monodisperse gold nanorods (AuNRs) with strong photothermal effect as the core.[103] In the following, we give a few examples of nanoreactors that use PNIPAM networks as active triggers.

Polystyrene (PS)-PNIPAM core-shell microgels have been used as nanoreactors for metal nanoparticles.[45,101] Figure 3 gives a schematic presentation of a core-shell particle consisting of a solid core and a thermosensitive network. Nanoparticles of noble metals such as Pt, Pd or Ru can be generated within the network. Using the reduction of 4-nitrophenol as a model reaction, it has been shown that the catalytic activity of immobilized metal nanoparticles can be tuned by the swelling and shrinking of the thermosensitive microgels: The particles are dispersed in water which will penetrate the network at room temperature. Raising the temperature above 32°C a volume transition takes place and the now hydrophobic PNIPAM-network will expel most of the water. The dependence of the catalytic activity of the embedded nanoparticles exhibits a distinct minimum at the transition temperature and not the usual monotonous Arrhenius dependence observed for freely suspended particles.[45] Up to now, this finding made by us and by others has been interpreted in terms of decreased diffusion coefficients in the shrunken network which slows down mass transport considerably. However, we have demonstrated recently that the change of the polarity of the network above the volume transition leads to a different reactant partitioning inside the network which is a decisive factor that enters explicitly into the rate equation.[67] The catalytic activity it is therefore controlled by the combination of both processes, diffusion and partitioning, through the definition of a nanoreactor shell permeability.[68] A full discussion of this model which constitutes the theory of a nanoreactor will be given in section 4 below.

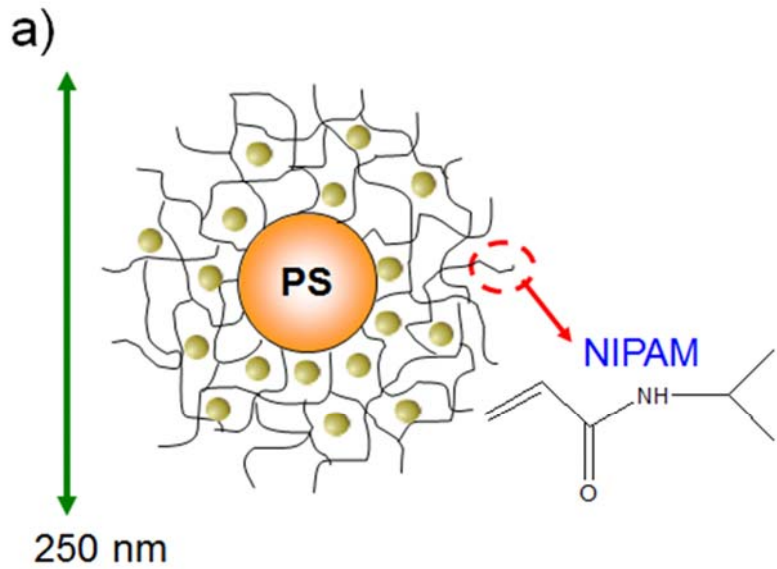

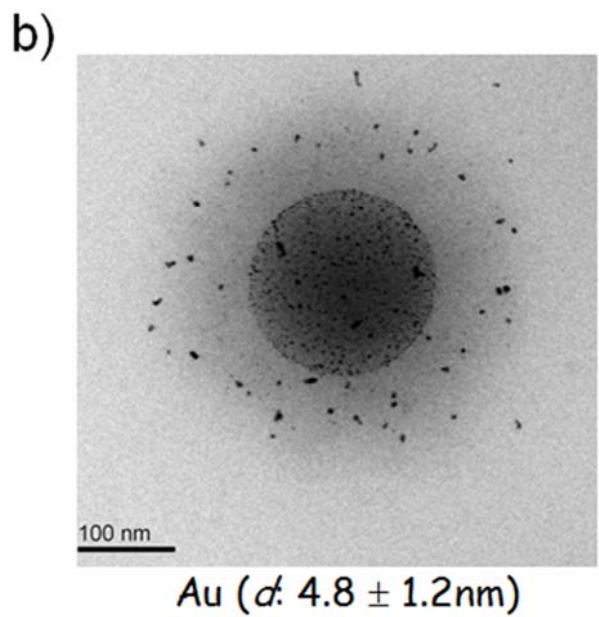

**Figure 3** Core-shell microgels as nanoreactors. a) The colloidal carrier particles consist of a solid core made from polystyrene onto which a highly crosslinked network of PNIPAM is grafted. b) Micrograph of a particle in aqueous phase taken by cryogenic transmission electron microscopy. Metal nanoparticles are generated within the shell. Their catalytic activity can be changed by the shrinking of the network with raising temperature.[101]

Shi et al. prepared PNIPAM-microgels and incorporated charged groups in order to generate sensitivity towards the pH as well.[46] Figure 4a and Figure 4b show the particles and the general concept in more detail. Gold nanoparticles were immobilized within the network. Using the reduction of nitrophenol, these workers could demonstrate the influence of the volume transition on the catalytic activity of the nanoparticles. They showed that the apparent rate constant was reduced by 81% of its value below the transition. Thus, this result clearly demonstrates the general power of the concept of a polymeric nanoreactor for the fine-tuning of the catalytic activity of metallic nanoparticles.

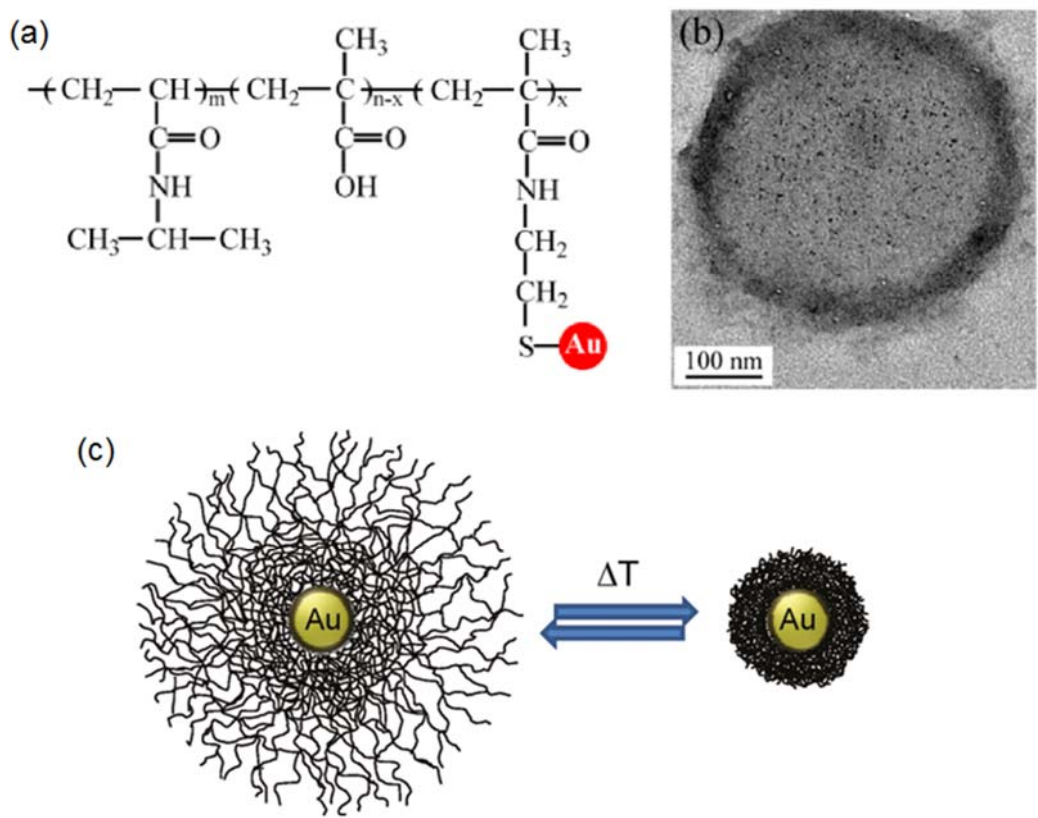

**Figure 4**. (a,b) Immobilization of gold nanoparticles in thermosensitive microgels. An Au-precursor is reduced within the network which carries thiol-groups. The swelling and shrinking can be triggered by external parameters (e.g., temperature or pH). The catalytic activity of the enclosed nanoparticles is strongly diminished in the vicinity of the temperature of the volume transition.[46] (c) Nanoreactors set up by single gold particles embedded in a thermosensitive PNIPAM network.[55] The hybrid particles are immersed in water that swells the PNIPAM shell at low temperature. Above the volume transition (at ca. 32°C) the network is in the shrunken state, and exhibits a larger diffusional resistance to the reactants compared to the open state at low temperature. Adapted from reference.[55]

Carregal-Romero *et al.*[55] have coated Au nanoparticles with a thermosensitive poly(N-isopropylacrylamide) (PNIPAM) microgel. Figure 4c shows the system in a schematic fashion. They used the reduction of hexacyanoferrate(III) by borohydride ions as a model reaction[104] (see below), and found that the catalysis of encapsulated Au nanoparticles can be controlled by the thermoresponsive shell. The rate constant measured by UV/VIS-spectroscopy exhibits a sharp drop around the temperature of the volume transition. This finding was modelled in terms of an decreased diffusion coefficient caused by the shrinking of the network, and the theory by Carregal-Romero et al.[55] provided the first theoretical modeling of a nanoreactor (cf. section 4).

Liu at al. presented an interesting extension of these concepts by using magnetic core particles covered by a PNIPAM-network in which Au-nanoparticles have been embedded.[60] The thermosensitive network can be used to tune the catalytic activity of the nanoparticles by changing the temperature while the magnetic core allows an easy separation of the hybrids from the solution after catalysis.

Wang et al. have synthesized thermo-responsive diblock copolymers poly(N-isopropylacrylamide-co-2,3-epithiopropyl methacrylate)-*block*-poly(poly(ethylene glycol) methyl ether methacrylate) P(NIPAM-co-ETMA)-b-P(PEGMA) by RAFT polymerization, which directly self-assembled into micelles with hydrophobic P(NIPAM-co-ETMA) blocks as the core in aqueous solution. The block copolymers containing sulfide groups were used as stabilizers for the *in-situ* preparation of Au nanoparticles. It is found that the constructed Au nanocomposites could act as a thermo-responsive catalyst and the catalytic activity could be adjusted by the thermoresponsive phase transition of the PNIPAM blocks.[105]

Nearly all nanoreactors are built into colloidal systems. The advantages of such colloidal particles as opposed to planar systems are at hand: Suspensions of colloids can provide a much larger surface than planar substrates and may thus lead to higher conversion per unit time. Also, a spherical system provides the smallest diffusional resistance possible. However, Koenig et al. demonstrated that planar polymer systems can work in the same way as colloidal systems. They prepared a planar thermoresponsive brush system, and immobilized Pt and Pd-nanoparticles within it.[58] Using the reduction of 4-nitrophenol, these workers could demonstrate that the apparent reaction rate has a non-monotonous dependence on $1/T$ as expected. It is worth mentioning in this regard that graphene-based systems[106] could provide interesting intermediate between planar and spherical systems, and this route remains to be explored in more detail.

Horecha et al. enclosed Ag nanoparticles attached to silica particles in microcapsules with a PNIPAM wall.[61] Surprisingly, they found that the rate constant increases with temperature even more for the encapsulated Ag-particles than for the free ones. Here the decreased reactant diffusion coefficient in the PNIPAM network was overcompensated by other effects, for example by a much larger reactant partitioning at higher temperature. This finding clearly demonstrates that the change of the rate constant with temperature achieved through nanoreactors needs more than the consideration of the reactant diffusion coefficient in the network (see section 4).

Yolk-shell type nanostructures in which metal nanoparticles are encapsulated in hollow porous shells have shown excellent stability of the metallic nanoparticle during different applications.[107,108] These systems can be used to tune the catalytic activity of the enclosed metal nanoparticle by a suitable architecture of the shell. To this purpose, yolk-shell systems with nanoparticles embedded in different shells have been developed, based on inorganic capsules,[109–113] carbon capsules[114] or polymer microcapsules.[36,57,61,115] The yolk–shell particles have the clear advantage that the encapsulated metal nanoparticle has a free surface, which is not blocked by any surface group compared to other architectures.

Thermosensitive Au-PNIPAM yolk-shell systems in which a single Au-nanoparticles is immobilized in a hollow shell of PNIPAM have been synthesized by us.[36] The selectivity and activity of this catalytic hybrid system can be tuned by temperature as is shown by the competitive reduction of the hydrophilic 4-nitrophenol and the hydrophobic nitrobenzene by borohydride. As shown in Figure 5, the more hydrophilic substrate 4-nitrophenol reacts at a higher rate at room temperature while the more hydrophobic nitrobenzene reacts at a much higher rate at elevated temperature. Again this result cannot be explained by reduced reactant diffusion in a denser gel but reactant partitioning must be invoked as well.

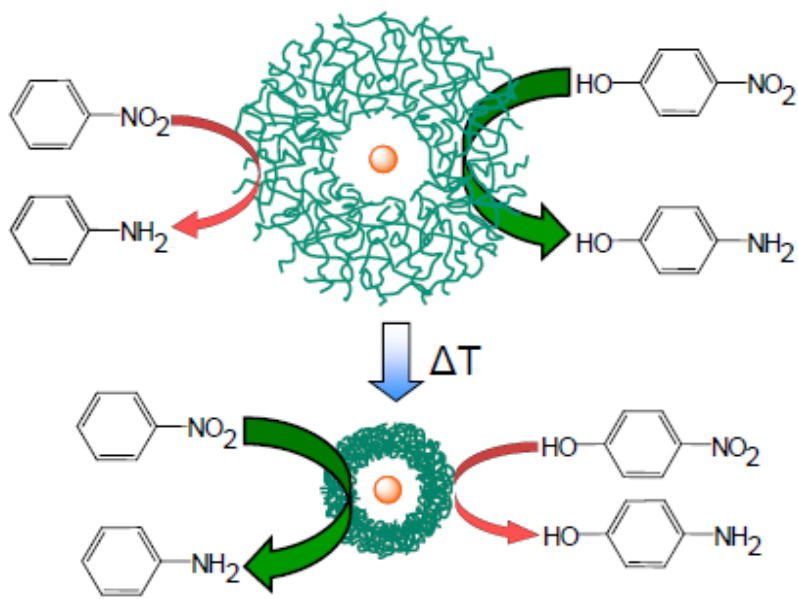

**Figure 5** Nanoreactor based on a thermosensitive yolk-shell system. A gold nanoparticle is enclosed in a hollow sphere made from a thermosensitive network, which changes its state by increasing the temperature: At room temperature the hydrophilic network is swollen and hydrophilic substrates as 4-nitrophenol will preferentially penetrate. Thus, this substrate will react at a higher rate than a hydrophobic substrate like nitrobenzene. Above the temperature of the volume transition the polarity and the degree of swelling change. Now the hydrophobic substrate nitrobenzene will penetrate preferably, and react at a higher rate.[36]

Chen et al.[57] have fabricated a nanoreactor with a highly intricate architecture as shown in Figure 6: Here the Au-nanoparticles are located inside of a hollow and porous silica particle. The external pore mouths of mesoporous silica hollow spheres are covered by a PNIPAM gel. The test of the catalytic activity of the Au-nanoparticles was performed using the reduction of 4-nitrophenol and a much decreased activity was found at 50°C as opposed to the one measured at room temperature.

Diffusion was considered as the key factor for the catalytic reaction, which can be controlled by the PNIPAM layer.

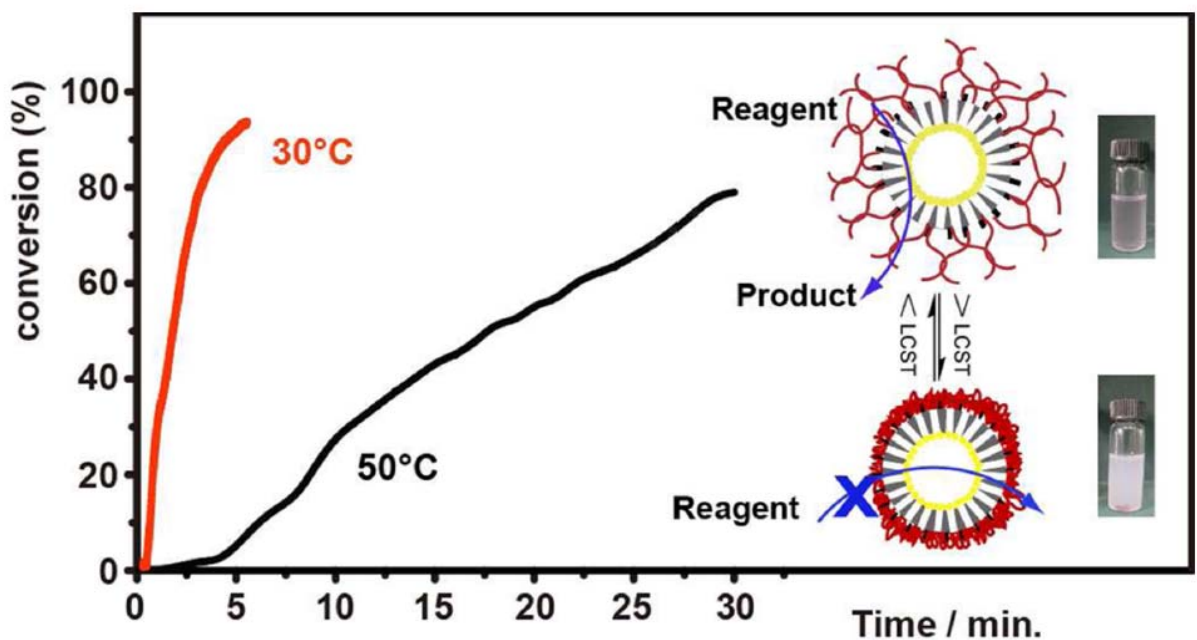

**Figure 6** Use of a thermosensitive PNIPAM-network to open and to close a nanoreactor: Au-nanoparticles are enclosed in a hollow porous silica capsule. The pores are covered by the network, the permeability of which can be changed by temperature.[57]

*Ligand-free nanoparticles as a reference for kinetic studies*: The above survey has shown that nanoreactors can profoundly modify the catalytic activity of nanoparticles. Ligand-free metal nanoparticles with uncapped metal surfaces can be regarded as a suitable reference system for a quantitative comparison of the activity of free nanoparticles as opposed to nanoparticles in carrier systems. These particles are generated in aqueous phase by laser ablation through an intense laser beam.[116–119] The nanoparticles are stabilized only by electrostatic interaction. These ligand-free nanoparticles can be used as a reference material for the above systems in mechanistic studies of catalytic activity:[119] There is no resistance for mass transport around these particles and no strongly attached ligand impedes their catalytic activity (see also the discussion in section 4). Thus, these systems are expected to exhibit the highest catalytic activity. The comparison with the catalytic activity of nanoparticles bound in carrier system may then give direct information on the influence of the carriers onto mass transport.

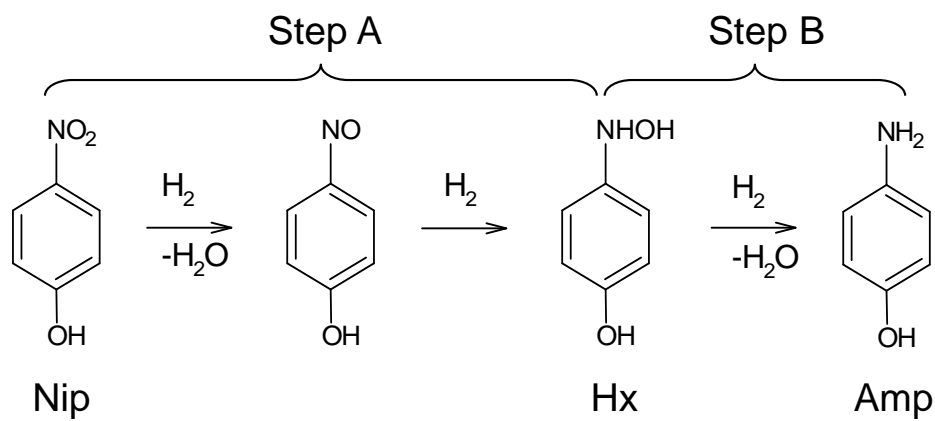

**Figure 7** Mechanism of the reduction of 4-nitrophenol by metallic nanoparticles along the direct route: In step A 4-nitrophenol (Nip) is first reduced to the nitrosophenol which is quickly converted to 4-hydroxylaminophenol (Hx) which is the first stable intermediate. Its reduction to the final product 4-aminophenol (Amp), takes place in the rate-determining step B. There is an adsorption/desorption equilibrium for all compounds in all steps. All reactions take place at the surface of the particles.

## 3. Model reactions for testing the catalytic activity of nanoparticles in solution

As mentioned above, a quantitative investigation of the catalytic activity of nanoparticles in various nanoreactors requires a chemical reaction that may be used to obtain precise kinetic data. Such model reactions can be defined in the following manner:[37]

1. It must be a simple and well-controlled chemical reaction that reacts one given substance A into a single product B. The reaction must be catalyzed by the nanoparticles under consideration and virtually no reaction must take place in absence of the particles. The order of the reaction must therefore be simple and most reactions that qualify as model reactions are of first order with respect to the main reactant.

2. A kinetic analysis of the reaction by simple and precise methods should be possible and the mechanism of the reaction including possible intermediate must be known precisely. Here optical techniques like UV-VIS spectroscopy are the preferred tools since they lead to a highly precise measurement of the conversion as the function of temperature.

3. The reaction should proceed under rather mild conditions and ambient pressure, preferably at room temperature in order to avoid all side reactions or a possible dissolution of the nanoparticles.

Evidently, these conditions are only fulfilled by surface reactions. Catalysis in this case is effected by active sites on the surface of the particles and not by any metallic species that has leached into the solution. Hence, the Heck- and the Suzuki-reaction are excluded from this list because all evidence points to homogeneous catalysis by small amounts of a dissolved species and not to a surface reaction.[80,120] Thus, testing a reaction for suitability must always include the dependence of the rate constants on the total surface of the particles in the system. Normalization of the rate constant to this surface is therefore the first requirement for a meaningful comparison of different systems.[14,67]

*Reduction of 4-nitrophenol by borohydride ions*

The reduction of 4-nitrophenol by borohydride ions in aqueous solution is by far the most used model reaction for testing the catalytic activity of metallic nanoparticles.[63–65,121] Its mechanism can be inferred from the well-studied reduction of nitrobenzene in solution. As early as 1898 F. Haber[122] demonstrated in a series of electrochemical experiments that there are two routes by which the nitro-compound is reduced to the respective amino-compound:[123,124] In the *direct route* (see Figure 7 and the review by Blaser[124]) the nitroarene (Nip) is first reduced in step A to the nitroso-compound which in a fast reaction is turned to the respective hydroxylamine (Hx). In step B which is rate-determining the hydroxylamine is reduced to the final product, namely the amine (Amp). In the *condensation route* the nitroso compound can react with the respective hydroxylamine to form azoxybenzene and subsequent reduction products. Recent investigations have clearly shown that the reduction proceeds only along the direct route if catalyzed by Au-nanoparticles and no traces of azoxybenzene or of the following products can be detected.[66,123,125,126] Hence, the mechanism of the reduction of 4-nitrophenol catalyzed by nanoparticles can safely be discussed in terms of the direct route shown in Figure 7.[127]

Nearly all investigations reported so far came to the conclusion that all steps of this reaction take place on the surface of the nanoparticles. This can be seen, e.g., from the fact that the apparent reaction rate scales with the total surface of the nanoparticles in solution.[14] Additional work has recently corroborated this result.[15,128,129] Moreover, the fact that the reaction proceeds only along the direct route lends further support for this finding: The concentration of the intermediates in solution is low because of the strong adsorption to the metal surface. Hence, the probability for a reaction between these intermediates in free solution is vanishingly small. Recently, Nigra et al.[130] came to conclusion that the catalysis of the reduction of Nip in presence of gold nanoparticles is done by a soluble Au-species. However, the reduction of nitrophenol is catalyzed also by other noble metals such as Pt[14,94,126], Ag[45,131,132], Pd,[26,133] Ru[15], and by nanoalloys.[133] The catalytic activity of these metals is of similar magnitude and it is safe to assume that the catalysis proceeds on the surface for all systems studied so far.

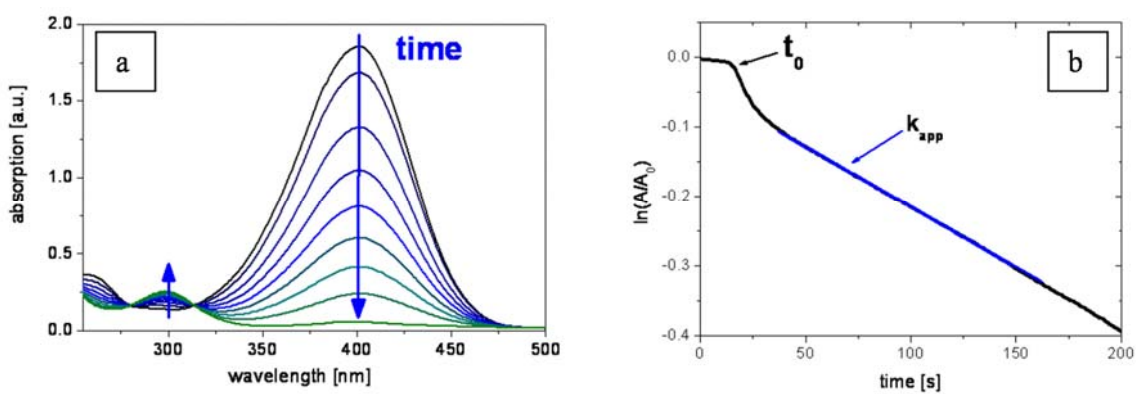

**Figure 8.** Reduction of nitrophenol by sodium borohydride as a model reaction for testing the catalytic activity of nanoparticles. (a) Absorption spectra of the solution as a function of time. The time between two consecutive curves is 2 minutes. Full conversion takes place in 20 minutes. The main peak at 400 nm (nitrophenolate ions) decreases as reaction time proceeds, whereas a second peak at 300 nm, referring to the product aminophenol, slowly increases (see arrows). Note the two isosbestic points are visible at 280 nm and 314 nm and indicate that only one product is formed during this reaction. (b) Typical time trace of the absorption of 4-nitrophenolate ions at 400 nm during their reduction. The reaction starts after an induction period $t_0$, The blue portion of the line corresponds to the linear section, from which $k_{app}$ can be accurately determined.[14,127]

The great advantage of this reaction is given by the fact that the consumption of 4-nitrophenol can be monitored easily by UV/VIS-spectroscopy. Figure 8a shows typical spectra measured at different conversion. The decay of the peak of the nitrophenolate ion at 400 nm can be followed precisely and the occurrence of several isosbestic points in the spectra points clearly to the conversion of the initial compound to a single final product. Figure 8b displays the temporal evolution of this peak in a plot of the logarithm of the ratio of the absorbance $A$ to the absorbance $A_0$ at $t = 0$. The concentration of the reducing agent sodium borohydride has been much higher so that a plot for a first order reaction can be used. After a delay time $t_0$ the reaction starts and the rate given by the tangent of this plot is larger at first. Finally, a stationary state results and the slope in this region has been taken for the apparent rate constant $k_{app}$.

The analysis of the details of the kinetics of the reaction can be done on the level of phenomenological rate equations as follows: First, as the reaction is fully surface reaction-controlled, the apparent rate constant $k_{app}$ is directly proportional to the total surface $S$ of all nanoparticles in solution:[67,127]

$$-\frac{dc_{Nip}}{dt} = k_{app} c_{Nip} = k_1 S c_{Nip} \qquad (1)$$

The resulting reduced rate constant $k_1$ has been evaluated by many workers in the field and will be discussed further below.[37] In a second step the full analysis can be done taking into account all the intermediates shown in Figure 7.[127] This can be done in terms of a Langmuir-Hinshelwood analysis. Here both reacting species are assumed to be adsorbed to the surface in order to react. Hence, for step A (Figure 8) we require the surface coverage of nitrophenol $\theta_{Nip}$ and of borohydride $\theta_{BH4}$ to formulate the rate equation which leads to

$$-\frac{dc_{Nip}}{dt} = k_{app}c_{Nip} = k_aS\theta_{Nip}\theta_{BH4} = \left(\frac{dc_{Hx}}{dt}\right)_{source} \quad (2)$$

While the adsorption/desorption equilibria of 4-nitrophenol is easy to model, the interaction of the borohydride ions with the metal surface is more complicated. Work done on catalytic decomposition of $BH_4^-$ ions by metal nanoparticles suggests the formation of a surface hydrogen species which reacts with the adsorbed nitroarene (see ref.[126] and further references therein). Admittedly, this problem is in need of further elucidation.

The central idea behind the modeling through eq. (2) is the competition of the two reacting species for the surface or active sites located at the surface. If the surface coverage of the first reactant is exceeding a certain value it will block the site for the second reactant and the rate of reaction will be slowed down concomitantly. The constant $k_A$ is the true kinetic constant of the first step of the reaction and its magnitude is measuring the catalytic activity of a given surface for this step.

The first stable intermediate 4-hydroxyaminophenol reacts in step B again on the surface. The rate equation is therefore

$$-\left(\frac{dc_{Hx}}{dt}\right)_{decay} = k_aS\theta_{Hx}\theta_{BH4} = \frac{dc_{amp}}{dt} \quad (3)$$

The surface coverage of the respective compounds may be calculated by the classical Langmuir isotherm

$$\theta_{Nip} = \frac{(K_{Nip}c_{Nip})^n}{1+(K_{Nip}c_{Nip})^n + K_{Hx}c_{Hx}+K_{BH4}c_{BH4}} \quad (4)$$

where $K_{Nip}$, $K_{Hx}$, and $K_{BH4}$ are the Langmuir adsorption constants of the respective compounds, and n is the Langmuir-Freundlich exponent. The surface coverages of the other reactants are given by analogous expressions. Thus, the kinetics of the reduction can be formulated in terms of two coupled differential equations:[127]

$$-\frac{dc_{Nip}}{dt} = k_a S \frac{(K_{Nip}c_{Nip})^n K_{BH4}c_{BH4}}{[1+(K_{Nip}c_{Nip})^n + K_{Hx}c_{Hx}+K_{BH4}c_{BH4}]^2} = \left(\frac{dc_{Hx}}{dt}\right)_{source} \quad (5)$$

and

$$\frac{dc_{Hx}}{dt} = \left(\frac{dc_{Hx}}{dt}\right)_{source} - \left(\frac{dc_{Hx}}{dt}\right)_{decay} = k_a S \frac{(K_{Nip}c_{Nip})^n K_{BH4}c_{BH4}}{[1+(K_{Nip}c_{Nip})^n + K_{Hx}c_{Hx}+K_{BH4}c_{BH4}]^2}$$

$$- k_b S \frac{K_{Hx}c_{Hx}K_{BH4}c_{BH4}}{[1+(K_{Nip}c_{Nip})^n + K_{Hx}c_{Hx}+K_{BH4}c_{BH4}]^2} \quad (6)$$

Considerable simplification can be achieved by the usual assumption of a stationary state through the condition

$$\frac{dc_{Hx}}{dt} = 0 \quad (7)$$

which leads to the following expression for the reaction rate:[127]

$$-\frac{dc_{Nip}}{dt} = k_a S \frac{(K_{Nip}c_{Nip})^n K_{BH4}c_{BH4}}{\left[1+(K_{Nip}c_{Nip})^n(1+\frac{k_a}{k_b})+K_{BH4}c_{BH4}\right]^2} = \frac{dc_{amp}}{dt} \quad (8)$$

Thus, in this stationary state the decay of the concentration of nitrophenol is matched exactly by the raise of the concentration of the final product 4-aminophenol. Hence, isosbestic points are seen under these conditions despite the fact that 3 compounds are involved. In the first non-stationary stage there can be no isosbestic points, of course (see ref.[127] for a detailed discussion of this problem).

Up to now, this model has been compared to experimental data for ligand-free nanoparticles[119] and for nanoparticles and –alloys bound to spherical polyelectrolyte brushes.[127,134] In all cases there was good agreement of theory and experiment. In particular, the two-step decay shown in the kinetic data (see Figure 8b) can be explained as stated above: At first 4-hydroxyaminophenol is generated which blocks the active sites on the surface of the nanoparticles since its reduction is the rate-determining step. After this initial phase a stationary state evolves in which the overall concentration of the intermediate 4-hydroxyaminophenol becomes a constant (see eqs. (7) and (8)). Evidently, the model possesses a number of adjustable parameters as the three adsorption constants as well as the two kinetic constants $k_A$ and $k_B$. However, the system of the two coupled differential equations is quite stiff and good estimates of all constants can be obtained (see the discussion of this point in ref.[127]). A separate analysis of the concentration of the intermediate 4-hydroxyaminophenol, however, would be clearly desirable. We note that the application of this kinetic model to responsive nanoreactors[36] and the consistent inclusion of shell permeability effects (as discussed in the theory section below) is work in progress.

It rests to explain the delay time $t_0$ seen in many studies of this reaction (see Figure 8b). Wunder et al. have presented evidence that this delay time may be traced back to a surface restructuring of the particles bound to spherical polyelectrolyte brushes.[66,126] First of all, it can be demonstrated that there is no strong diffusional resistance for these systems that could possibly generate delay times of the order of minutes. On the other hand, $t_0$ can be related to the degree of surface coverage $\theta_{Nip}$ of the nanoparticles by the 4-nitrophenol as shown in Figures 7 and 8 of reference[66]. The reciprocal delay time can be taken as a measure for the rate with which the surface is restructured. It was found that $1/t_0$ scales linearly with $\theta_{Nip}{}^2$ which provides strong evidence for surface restructuring. Recently, Menumerov et al.[135] advanced the hypothesis that the delay time is due to the reaction of borohydride with dissolved oxygen. This problem had been discussed in earlier papers[14] and in ref.[136]. It was demonstrated in the latter reference that $t_0$ is independent of the initial concentration of borohydride ions which is difficult to explain by the model of Menumerov et al. Further work is needed, however, to elucidate this problem in closer detail.

*Reduction of hexacyanoferrate (III) ions by borohydride ions*

Another electron transfer reaction that has been used as a model reaction to study catalysis by metallic nanoparticles is the reduction of hexacyanoferrate (III) by borohydride ions.[55] Figure 9 displays schematically the mechanism of the reaction: Electrons are injected into the metal nanoparticles and the electron transfer to the hexacyanoferrate(III)-ions takes place at the surface. This reaction in aqueous solution can be written as:[55,106,137]

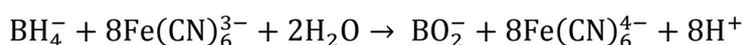

$$BH_4^- + 8Fe(CN)_6^{3-} + 2H_2O \rightarrow BO_2^- + 8Fe(CN)_6^{4-} + 8H^+$$

As shown by Figure 9, the reaction can be accurately followed by UV/VIS-spectroscopy and the decay rate can be modelled by a kinetics of first order (see inset of Figure 9). The redox potential corresponding to the reaction, $E^0(Fe(CN)_6{}^{3-}/ Fe(CN)_6{}^{4-})$ is +0.44 V versus a normal hydrogen electrode (NHE). Since the standard reduction potential for borate ion is $E^0 = -1.24$ V vs. NHE, the free energy change associated with the reaction is very pronounced. However, in the absence of a catalyst, there is a slow reaction (half-life around 5000 s; see the discussion in ref.[37] and further citations given therein).

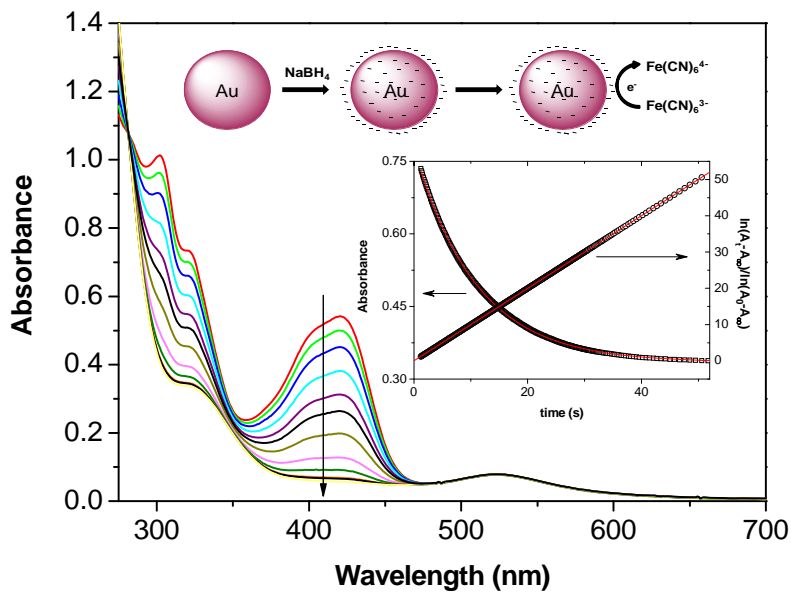

**Figure 9.** Spectral evolution of a mixture of hexacyanoferrate (III) and Au nanoparticles upon borohydride addition. The inset shows the absorbance at 420 nm as the function of time during the reduction of $Fe(CN)_6^{3-}$. It demonstrates that the reaction is of first order. The inset on top of the graph illustrates the proposed mechanism: Borohydride ions inject electrons into the nanoparticles which reduce the $Fe(CN)_6^{3-}$ ions, much in the way of an electrochemical reaction at a macroscopic electrode.[55,104]

The work of Carregal-Romero et al.[55,104] has clearly demonstrated that gold nanoparticles are efficient catalysts for the reduction of hexacyanoferrate(III)-ions. Therefore, this reaction can be regarded as a model reaction for electrochemistry at the nanoscale. Related electron injection processes were reported in the past for gold and silver colloids, which demonstrated this ability of nanoparticles to function as nanoelectrodes.[139,140] Thus, once the nanoparticle surface is charged by the addition of a reducing agent, the stored electrons can be discharged if an electron acceptor is introduced into the system. Using this model reaction, Carregal-Romero et al.[55] were able to show that the reactivity of a gold nanoparticles in a thermosensitive PNIPAM network can be tuned by temperature as indicated schematically in Figure 4c: At low temperature the PNIPAM network is swollen and the ions can diffuse freely to the enclosed nanoparticle. At high temperature the network shrinks and the catalytic activity is decreasing considerably (see Figure 8 of ref.[55] and the discussion given therein). The authors explain this finding by a marked decrease of ion diffusion in the shrunken state. The same authors also developed a theoretical model for a diffusion-limited reaction in the temperature-sensitive shell that is consistent with the experimental data.[55] This model (see SI of ref.[55]) and its extension presented by us will be the subject of a detailed discussion in the next section.

### 4. Nanoreactors: Theory and modeling

*Yolk-shell systems*

We now summarize our recent efforts towards a theory for the reactions in nanoreactors. As discussed in section 1, nanoreactors consist of metal nanoparticles embedded in a polymeric or inorganic structure. We first treat the most simplistic case, namely a yolk-shell system in which a single metal nanoparticle is embedded in a hollow sphere of a thermosensitive network.[36,67,68] In general, the chemical reactivity of these nanoparticles will have two contributions, namely a part related to the surface reaction and a part related to mass transport by diffusion. All the aforementioned model reactions deal with surface-catalyzed bimolecular reactions. In our recent work, we showed that the total catalytic rate in diffusion-influenced bimolecular reactions is given by[68]

$$k_{tot} = \frac{1}{2}\left[\frac{k_{D_A}k_{D_B}}{k_R} + k_{D_A} + k_{D_B} - \sqrt{\left(\frac{k_{D_A}k_{D_B}}{k_R} + k_{D_A} + k_{D_B}\right)^2 - 4k_{D_A}k_{D_B}}\right] \quad (9)$$

where $k_R$ is the surface-controlled rate, and $k_{D_A}(\mathcal{P}_A)$ and $k_{D_B}(\mathcal{P}_B)$ are the diffusion-controlled rates of the reactants A and B, which explicitly depend on the shell permeability, $\mathcal{P}$, defined as

$$\mathcal{P} = D_g \mathcal{K} \quad (10)$$

In the previous equation, $D_g$ represents the reactant diffusivity in the shell, and $\mathcal{K}$ is the partitioning of reactants, i.e., the ratio between their concentrations inside and outside the shell

$$\mathcal{K} = \frac{c_g}{c_0} = e^{-\Delta G_T} \quad (11)$$

Equation (9) shows that the total rate in bimolecular reactions depends in a nontrivial way on the reactant diffusion and surface reaction rates. In the case of a limiting reactant, e.g., if one substrate is very dilute or diffuses very slowly, a bimolecular reaction can be treated as pseudo-unimolecular and eq. (9) simplifies to

$$k_{tot}^{-1} = k_D^{-1} + k_R^{-1} \quad (12)$$

being $k_D$ the diffusion rate of the limiting reactant. Thus, if the reactivity of the nanoparticles under consideration has been determined for the free particles in terms of a rate constant $k_R$, the rate constant measured for particles embedded in a given carrier $k_{tot}$ may then give quantitative information on the influence of diffusional transport. We note that so far we have only considered the low reactant adsorption limit and neglected intermediate reaction effects on the nanoparticle for $k_R$ as discussed in section 3. The permeability-dependent diffusion rate can be modelled in terms of the Debye-Smoluchowski equation:

$$k_D = 4\pi c_0 \left[\int_{R_0}^{\infty} \frac{1}{\mathcal{P}(r)r^2} dr\right]^{-1} \quad (13)$$

The apparent rate constant, $k_{app}$, for the pseudo-first order reaction given by eq. (1) is related to the pseudo-unimolecular (microscopic) total rate, $k_{tot}$, that appears in eq. (12) by

$$k_{app} = \frac{c_{nano}}{c_0} k_{tot} \qquad (14)$$

where $c_{nano}$ represents the concentration of nanoreactors in solution and $c_0$ is the bulk concentration of the limiting reactant.

Most of the studies in the literature assume to be dealing with pseudo-unimolecular reactions, usually by working with one of the reactants in large concentration excess. This assumption is valid for reactions in bulk solution. However, as we recently pointed out[68], care should be taken in the case of nanoreactors. The reason is that in these systems it is not the bulk concentration what matters but the reactant concentration at the nanoparticle surface where the reaction can take place. The latter is strongly influenced by the shell permeability which can strongly differ from the reference bulk value.

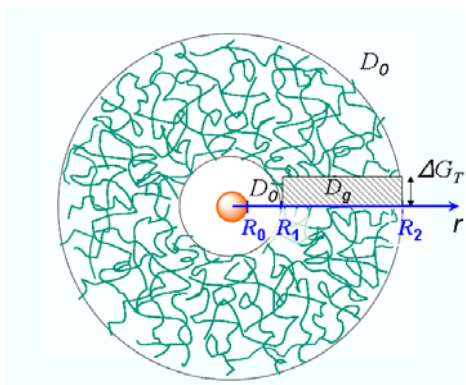

**Figure 10** Scheme of a yolk-shell nanoreactor: A single metal nanoparticle is embedded in a hollow shell composed of e.g. a polymeric network. The overall reaction rate $k_{tot}$ is composed of a term $k_R$ related to the surface reaction on the metal nanoparticles whereas the mass transport by diffusion is characterized by $k_D$ (see eq. (12)). Diffusion in free solution is described by the diffusion coefficient $D_0$ whereas diffusion through the network is characterized through $D_g$. The transfer free energy $\Delta G_T$ describes the change of free energy of a given reactant when going from the free solution into the network.

Figure 10 displays the underlying model in the spirit of the Debye-Smoluchowski rate theory: The transport outside of the nanoreactor is determined by the diffusion constant $D_0$ which in general will be larger than the diffusion constant $D_g$ within the shell characterized by the outer radius $R_2$ and the inner radius $R_1$. The radius of the nanoparticles is $R_0$. The quantity $G(r)$ is the Gibbs free energy of solvation of the reactant in its environment. The reference state is the free solution without any polymers. Hence, $G(r)$ denotes the gain or loss in free energy when a reactant molecule is transferred into the shell. It thus determines the Nernstian distribution of the reactant molecule between the network and the outside reservoir. For inhomogeneous layers this free energy may depend on the radial distance $r$, for most practical purposes it can be approximated by its mean value $\Delta G_T$, the mean transfer free energy for a reactant from bulk into the shell. Moreover, eq. (13) can be considerably simplified by the assumption that $D_g \ll D_0$ so that mass transport through the free solution is fast enough to be disregarded (see ref.[67] for a complete discussion of these points). Under these conditions eq. (13) leads to a simple expression[37,65] which is reminiscent of the classical Arrhenius equation:[36,67,68]

$$k_D \simeq 4\pi c_0 R_1 D_g \exp\left(-\frac{\Delta G_T}{k_B T}\right) = 4\pi c_0 R_1 \mathcal{P} \qquad (15)$$

Hence, the diffusion-controlled rate is proportional to the shell permeability $\mathcal{P}$. The diffusion coefficient enters linearly into the shell permeability, whereas its dependence on $\Delta G_T$ is exponential. Evidently, small changes of the latter quantity will exert a profound influence on $k_D$. Moreover, this formula clearly points towards a simple possible mechanism to explain why the reaction rate in nanoreactors can be sometimes higher with respect to free nanoparticles. In practice, a lower diffusion constant in the polymeric shell can be counterbalanced by a decrease in the transfer free energy.[67,68] Importantly, it should be pointed out that this free energy is to be measured with respect to the free energy of a reactant in the bulk solution. Thus, changes in the solvent can also influence the reaction rate.

We have shown in ref.[67,68] that the eq. (15) can be combined with a thermodynamic two-state model[55] describing temperature or pH induced changes of the polymer shell. In this way, trends in the activity with respect to these external triggers for active nanocarriers can also be rationalized.[67] Figure 11 displays a semi-quantitative comparison of theory and experiment.[67] Here the reduction of a hydrophilic molecule (blue lines) and a hydrophobic molecule (red lines) is catalyzed by a gold nanoparticle embedded in a PNIPAM network as depicted schematically in Figure 10. The external trigger is the temperature $T$, scaled by the temperature of the volume transition $T_c$ of the PNIPAM network (the kinetic rate constants have also been normalized to their maximum value found within the temperature range considered). The upper panel displays the calculated kinetic data whereas the lower panel gives the experimental data taken from refs.[36,55] (see also the discussion of Figure 5 above). Below the transition temperature the hydrophilic PNIPAM network is more permeable for the hydrophilic reactant leading to a higher reaction rate as compared to the one measured for the hydrophobic substrate. Above $T_c$ this relation is inversed. As already discussed above the main reason for this switching of the reactivity is the transfer free energy $\Delta G_T$ which enters exponentially into the rate eq. (15) whereas $D_g$, the diffusion coefficient of the substrate through the network enters only linearly. Evidently, an increased diffusional resistance of a shrunken network can be counterbalanced by $\Delta G_T$ as discussed above.

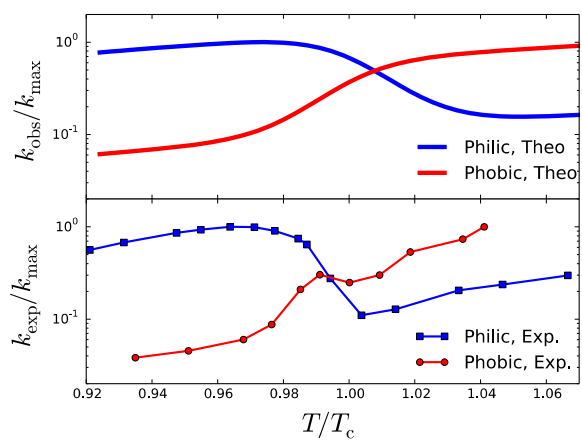

**Figure 11** Modelling of reactivity of metal nanoparticles in thermosensitive PNIPAM-based nanoreactors. Comparison of theory (eq. (15); upper panel) with experiments (taken from refs.[36,55]; lower panel) for a hydrophilic reactant (blue lines) and a hydrophobic reactant (red lines).[67]

*Nanoparticles embedded at random in a thermosensitive network*
Yolk-shell systems as depicted in Figure 10 present certainly the simplest nanoreactors for which a theoretical model can be set up. On the other hand, systems in which many nanoparticles are embedded at random in a thermosensitive network are much easier to synthesize. Hence, the composite particles shown in Figure 3 and Figure 4b are rather easily accessible and present stable nanoreactors for catalytic reactions in the aqueous phase. Recently, Galanti et al. extended the model of ref.[67] to describe unimolecular reactions in core-shell particles with multiple embedded nanoparticles[141]. Figure 12 shows these systems and their description in a schematic fashion:

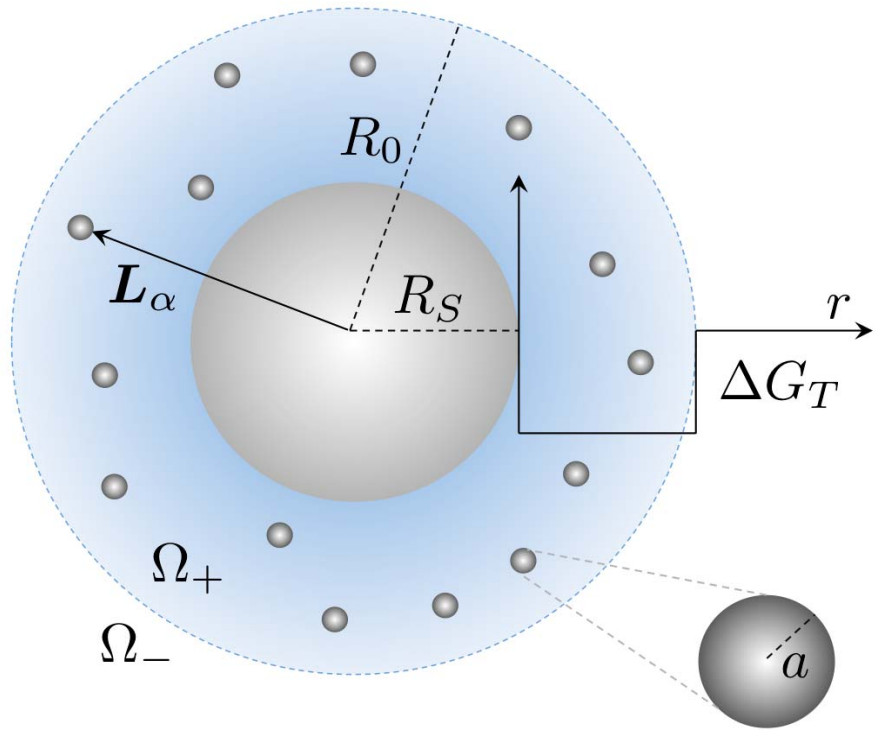

Figure 12. Scheme of a core-shell nanoreactor consisting of a solid core and a thermosensitive shell into which nanoparticles are embedded at random at positions $L_a$. The radius of the core is $R_S$ while the radius of the entire particle is $R_0$. $\Delta G_T$ is the difference between the free enthalpies of the solute inside and outside of the network. If $\Delta G_T < 0$, the solute is attracted to the network and its concentration within the network will be enhanced. In the opposite case, namely $\Delta G_T > 0$, the solute will be repelled from the network and its concentration within the shell diminished.[141]

Here $N$ catalytic nanoparticles are embedded at random in the shell of a core-shell nanoreactor as already shown in Figure. 3. The transfer free enthalpy $\Delta G_T$ can be smaller or greater than zero and will thus enhance or diminish the concentration of the solute reactant within the network, respectively. The solute reactant will enter into the network from the bulk phase and react at the surface of the $N$ catalytic nanoparticles. Obviously, modeling such a nanoreactor is more complicated because one must take into account the competition of $N$ reaction centers located at certain positions $L_\alpha$. The model by Galanti et al.[141] is fully general and treats the interplay between diffusional control of the reaction and control by the chemical reaction at the surface. The exact solution of the many-body problem detailed in Ref.[141] is rather involved, however the model leads to simple predictions for limiting cases. Thus, for a nanoreactor with a thick shell and a dilute distribution of $N$ nanoparticles, the overall rate constant $k$ in case of diffusion control follows as

$$\frac{k}{k_S} = \frac{N\varepsilon\xi e^{-\beta \Delta G_T}}{1+N\varepsilon\xi e^{-\beta \Delta G_T}} \qquad (16)$$

where $\beta = 1/kT$, $\varepsilon = a/R_0 \ll 1$ is the ratio between the radius of the core and the radius of the entire particle, while $\xi = D_i/D_0$ denotes the ratio of the diffusion coefficients of the solute inside and outside the network, respectively. The quantity $k_S = 4\pi D_0 R_0$ is the Smoluchowski rate for the whole nanoreactor that describes the total flux of solute molecules to a fully absorbing sphere of the same size as the nanoreactor, radius $R_0$, in units of the solute concentration.

Equation (16) shows that the rate $k$ is described as a function of the number of nanoparticles $N$ by a Langmuir-type expression. This finding can be easily traced back to the competition of an increasing number of nanoparticles for the substrate molecules within the network. For a single nanoparticle, that is $N = 1$, and a free enthalpy of transfer not too attractive, eq. (16) reduces to eq. (13) describing a single nanoparticle at the center of a spherical nanoreactor.

Obviously, there must be an optimal number $N_f$ of nanoparticles for a given geometry of the nanoreactor and a given level of *efficiency* of the nanoreactor. To discuss this point further, it is instructive to define the efficiency factor as $f = k/k_S$, that is, gauging the overall rate with respect to the maximum achievable rate $k_S$. From eq. (16) one is thus led to

$$N_f = \frac{f}{1-f} \frac{R_0}{a} \left(\frac{D_0}{D_i}\right) e^{\beta \Delta G_T} \qquad (17)$$

Thus, for a fixed efficiency $f$, the number $N_f$ depends exponentially on $\Delta G_T$. If we assume for example a reasonable set of parameters, namely $D_i/D_0 = 0.2$, $a/R_0 = 0.1$ and $\beta \Delta G_T = 0.1$. Then a number $N_f = 184$ is required to achieve an efficiency of 50%. For $\beta \Delta G_T = 0.2$, only 68 nanoparticles would be needed. This result demonstrates clearly the importance of $\Delta G_T$ for the efficiency of a core-shell nanoreactor. It also underscores the importance of the diffusivity of the substrate within the network embodied in $D_i$. Hence, unconstrained mobility of all reactants within the network improves the efficiency (less nanocatalysts required to get the same value of $f$) while sticky interactions via e.g. hydrogen bonding would slow it down.

For the limiting case of reaction-control, the model leads to

$$k \simeq N k^* e^{-\beta \Delta G_T} + \mathcal{O}[(k^*/k_S^+)^2] \qquad (18)$$

where $k^*$ is the turnover rate of the substrate at the surface of the nanoparticles. The expression can be again easily interpreted: The average concentration of the substrate within the network is increased by the factor $\exp[-\beta \Delta G_T]$. Thus, compared to the reaction of free nanoparticles in solution, the reaction rate is increased (or decreased) by the respective raise of the local concentration of the substrate. Hence, the model in Ref.[141] interpolates between reaction- and diffusion-control.

The above analytical treatment breaks down for efficiencies approaching unity of unity or values of the core radius $a$ approaching the overall size $R_0$. In this case the nanoparticles within the network do not act independently and diffusive interactions between them can no longer be neglected. Then the full expressions derived in Ref.[141] must be employed. The theory of Galanti et al. hence provides a comprehensive modeling framework for assessing the interplay of the various parameters governing the efficiency of nanoreactors.

**Conclusion**

The entire discussion has demonstrated that thermosensitive networks from PNIPAM with varying architecture can be used as active carrier systems for nanoparticles. Enclosed in such a nanoreactor, the catalytic activity of the nanoparticles can be adjusted by external triggers. Model reactions as discussed in section 3 allow us to obtain precise kinetic data that may be used to get quantitative information on these nanoreactors. Section 4 has shown that a theory of these nanoreactors that can account for the trends observed so far within a unified and thermodynamically consistent description is now available, and may be extended and used to interpret further experimental data. Work along these lines is under way.

**Corresponding Author** Matthias.ballauff@helmholtz-berlin.de

ACKNOWLEDGMENT: S.A-U. acknowledges financial support from the Beijing Municipal Government Innovation Center for Soft Matter Science and Engineering, as well as the Humboldt Foundation via a Postdoctoral Research Fellowship. J.D. and R.R. acknowledge funding by the ERC (European Research Council) Consolidator Grant with project number 646659–NANOREACTOR. F.P. acknowledges support from the French CNRS under the PICS scheme.

All authors acknowledge useful discussions with Daniel Besold, Won Kyu Kim, and Matej Kanduc.